\documentclass{appolb}
\usepackage{epsfig}
\def\beq{\begin{equation}}
\def\eeqno#1{\label{#1}\end{equation}}

\def\rarrow{\rightarrow }

\def\dleft{\rlap{{\it D}}\raise 8pt
\hbox{$\scriptscriptstyle\Leftarrow$}}
\def\dright{\rlap{{\it
D}}\raise 8pt\hbox{$\scriptscriptstyle\Rightarrow$}}
\def\DD{\dleft\dright}

\def\sk{S_k}

\def\az{a_{0}}
\def\azs{a_{0}^2}

\def\l0{\ell_{0}}

\def\rar{\rightarrow}

\def\a{\alpha}
\def\b{\beta}
\def\l{\lambda}

\def\c{\gamma}
\def\f{\phi}

\def\k{\kappa}
\def\r{\rho}
\def\m{\mu}
\def\n{\nu}

\def\AA{\textbf{A}}
\def\Ao{\mathcal{A}(\o)}

\def\o{\omega}

\def\D{\Delta}

\def\d{\delta}

\def\a{\alpha}
\def\xlimin{{x\rarrow\infty \atop{\raise 1pt\hbox to 30pt
{\rightarrowfill}}}}
\def\limlim#1#2{{#1\rarrow #2 \atop{\raise 1pt\hbox to 30pt
{\rightarrowfill}}}}

\def\vr{{\bf r}}

\def\vF{{\bf F}}
\def\vv{{\bf v}}
\def\vq{{\bf q}}

\def\va{{\bf a}}
\def\vA{{\bf A}}
\def\vf{{\bf f}}
\def\vF{{\bf F}}

\def\hao{\hat {\bf  a}(\o)}

\def\haNo{\hat {\bf  a}_N(\o)}
\def\hfo{\hat {\bf f}(\o)}

\def\grad{\vec\nabla}
\def\div{\vec \nabla\cdot}
\def\gf{\grad\phi}

\def\hao{\hat {\bf  a}(\o)}

\def\haNo{\hat {\bf  a}_N(\o)}
\def\hfo{\hat {\bf f}(\o)}

\def\Ao{\mathcal{A}(\o)}

\def\gmn{g_{\m\n}}
\def\Gmn{g^{\mu \nu}}

\def\hGmn{\hat \Gmn}

\def\hgmn{\hat g_{\m\n}}
\def\m{\mu}
\def\a{\alpha}
\def\b{\beta}
\def\c{\gamma}
\def\d{\delta}
\def\C{\Gamma}
\def\hC{\hat\Gamma}
\def\mt{\tilde\mu}
\def\n{\nu}
\def\Up{\Upsilon}
\def\ten#1#2{^{#1}_{#2}}

\def\emn{\eta_{\m\n}}
\def\fh{\hat\f}
\def\D{\Delta}
\def\fpg{4\pi G}

\def\ft{\tilde\f}
\def\rh{\hat\rho}
\def\fb{\bar\f}
\def\gfb{\grad\fb}
\def\lesssim{<\sim}
\def\baz{\bar a_0}
\begin{document}
\title{MOND--particularly as
modified inertia%
\thanks{Presented at XXXV International Conference of Theoretical Physics
MATTER TO THE DEEPEST: Recent Developments in Physics of Fundamental Interactions, Ustro\'n, Poland September 2011
}%
}
\author{Mordehai Milgrom
\address{Department of Particle Physics and Astrophysics, Weizmann Institute}}
\maketitle
\begin{abstract}
After a succinct review of the MOND paradigm--with its phenomenology, and its various underlying theories--I concentrate on so called modified inertia (MI) formulations of MOND, which have so far received only little attention. These share with all MOND theories the salient MOND predictions, such as asymptotically flat rotation curves, and the universal mass-asymptotic-speed relation. My emphasis here is, however, on the fact that MI theories can differ substantially from their ``modified-gravity'' (MG) kin in predicting other phenomena.
Because MI theories are non local in time, MOND effects depend on the full trajectory of a system, not only on its instantaneous state, as in MG theories. This may lead to rather different predictions for, e.g., the external-field effect (EFE):
A subsystem, such as a globular cluster or a dwarf galaxy, moving in the field of a mother galaxy, or a galaxy in a cluster, may be subject to an EFE that depends on the accelerations all along its orbit, not only on the instantaneous value. And, it is even possible to construct MI theories with practically no EFE.
Other predictions that may differ are also discussed.
Since we do not yet have a full fledged, modified-inertia formulation,  simple, heuristic models have been used to demonstrate these points.

\end{abstract}

\section{MOND and its phenomenology}
First I review somewhat telegraphically, the present status of the MOND paradigm. A more detailed recent review can be found in \cite{milgrom11}.
MOND was propounded almost 30 years ago \cite{milgrom83} as an alternative to dark matter (DM). A MOND theory is a theory of dynamics, replacing Newtonian dynamics in the nonrelativistic (NR) regime, and general relativity (GR) in the relativistic regime. It introduces into physics a new constant
$\az$, with the dimensions of acceleration. This plays similar formal roles to $c$ in relativity, or to $\hbar$ in quantum theory. In the formal limit $\az\rar 0$ a MOND theory is required to reduce to standard dynamics (similar to the correspondence principle in QM). The opposite, MOND limit, is defined by $\az\rar \infty, ~~G\rar 0$, with $G\az$ fixed, in which limit we require scale invariance, namely invariance of the NR theory under $(t,\vr)\rar\l(t,\vr)$.
For example, just on dimensional grounds, the acceleration of a test particle at a distance $R$ from a point mass $M$ has to be given by
$a= (MG/ R^2)f(MG/ R^2\az)$, where $f(x)$ depends on the theory, and possibly on the type of orbit involved. The correspondence principle requires that $f(x\rar \infty)\rar 1$. Scale invariance in the MOND limit requires that $ a\approx (MG\az)^{1/2}/R$, for $a\ll \az$. The function $f(x)$ is an example of the so called interpolating function. Such (possibly different) functions may appear in other contexts of MOND, and they are analogous to the Lorentz factor in relativity, or to the black-body function in quantum theory.

\subsection{ Independent
Kepler-like laws in galaxies }
A MOND theory should enable us to calculate the full dynamics of individual systems from only their baryonic mass distribution. Also, a set of general laws that govern dynamics in galactic systems can be derive from essentially only the above basic tenets of MOND, as Kepler's laws follow from Newtonian dynamics. They are independent laws, in the sense that no subset of them follow from the others in the context of Newtonian dynamics with DM. Here are some of these:
\par\noindent
1.  Asymptotic constancy of the orbital velocity around a bounded mass, $M$: $V(r)\rar
V_{\infty}$.
\par\noindent
2. The asymptotic-velocity-mass relation:  $V_{\infty}^4=MG\az$.
\par\noindent
3. A mass-velocity-dispersion relation, $\sigma^4\sim MG\az$ , for spherical systems (``isothermal'' spheres).
\par\noindent
4. The mass discrepancy in disc galaxies appears always where   $V^2/R\approx\az$.
\par\noindent
5. Quasi-isothermal spheres have mean surface densities
$\bar\Sigma\lesssim \az/G$.
\par\noindent
6. The central surface density of ``dark halos'' is $\approx \az/
2\pi G$.
\par\noindent
7. Disc galaxies have a disc AND a
spherical ``DM'' components.
\par\noindent
8. Scale invariance and the pivotal role of $\az$ lead to nonlinearity, which leads to an external-field effect (EFE)(\cite{milgrom83}, and subsequent discussions).

\begin{figure}
\begin{tabular}{lll}
\includegraphics[width=0.3\columnwidth]{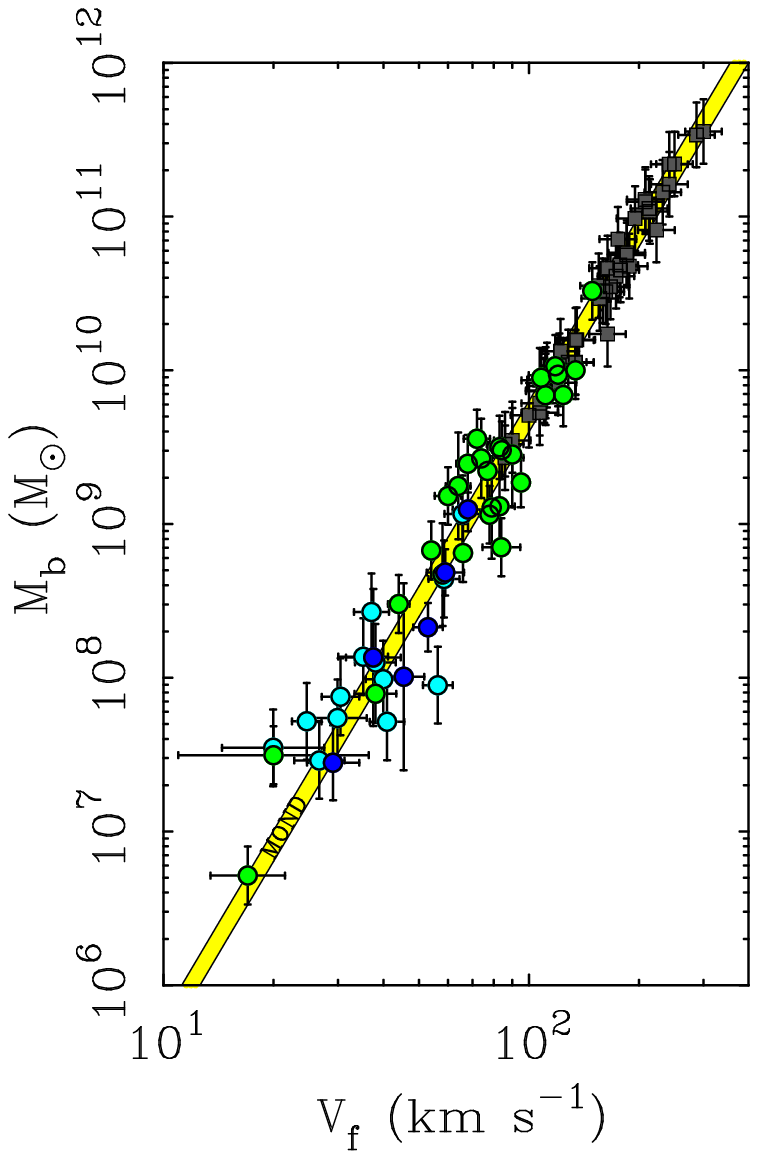}
\includegraphics[width=0.4\columnwidth]{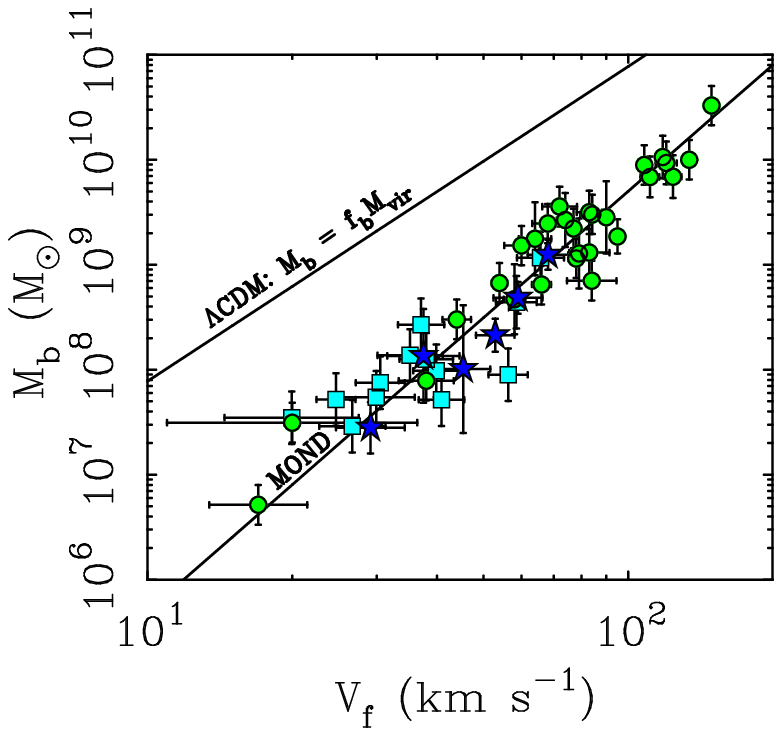}
\includegraphics[width=0.2\columnwidth]{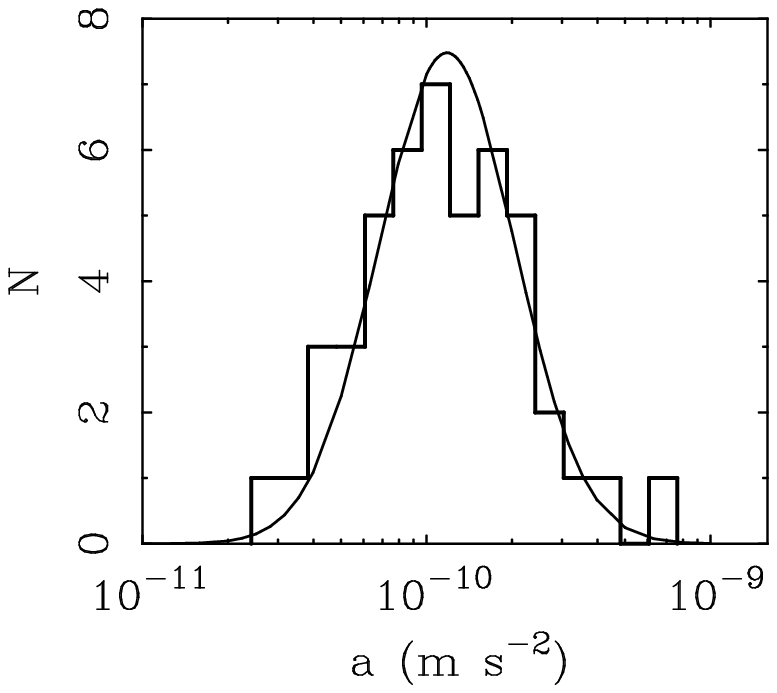}
\end{tabular}
\caption{Galaxy baryonic mass plotted against the asymptotic rotation speed. Left: A large sample of disc galaxies of all types. The yellow band is the MOND prediction (for a small range of allowed $\az$ values) \cite{mcgaugh11}. Middle: the same relation tested for a sample of gas-rich galaxies, for which the baryonic mass is insensitive to adopted $M/L$ values \cite{mcgaugh11a}. Right: distribution of $V^4/MG$ for the latter sample, compared with that expected from measurement errors alone; showing that the observed scattering is consistent with no intrinsic scattering in the observed relation.} \label{fig1}
\end{figure}

These predicted laws, inasmuch as they have been tested, were indeed found to hold in the observed galaxy dynamics. The many seemingly unrelated appearances of $\az$, in regularities that underlie galactic dynamics, are established facts, and remind one of the many appearances of $\hbar$ in disparate quantum phenomena. This is explained only within the unifying framework of MOND.
As an example, Figure \ref{fig1} shows the results of tests of the predicted asymptotic-velocity-mass relation (law 2 above).

\subsection{Rotation Curves of Disc
Galaxies}
\begin{figure}
\begin{tabular}{lll}
\includegraphics[width=0.36\columnwidth]{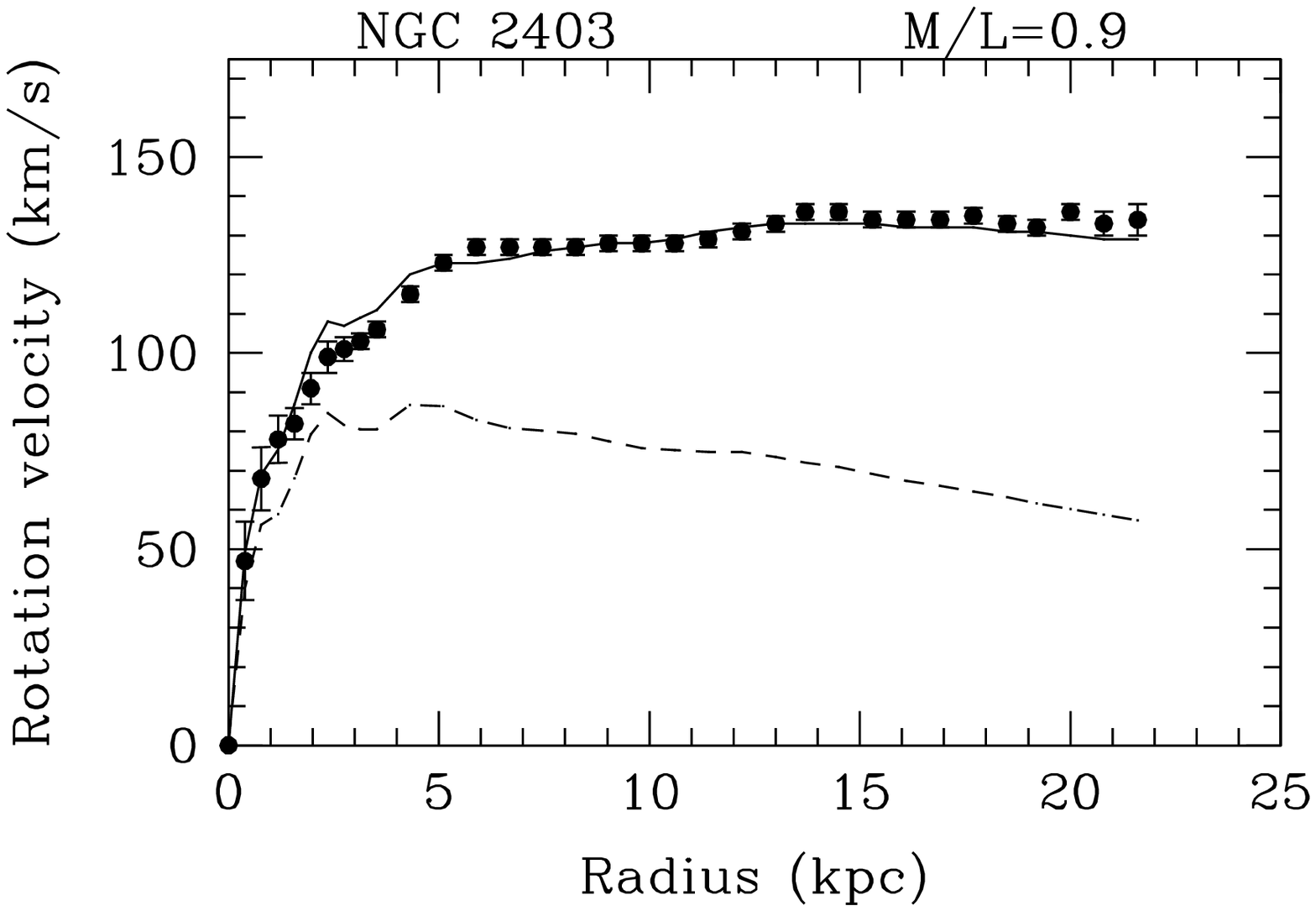}&
\includegraphics[width=0.32\columnwidth]{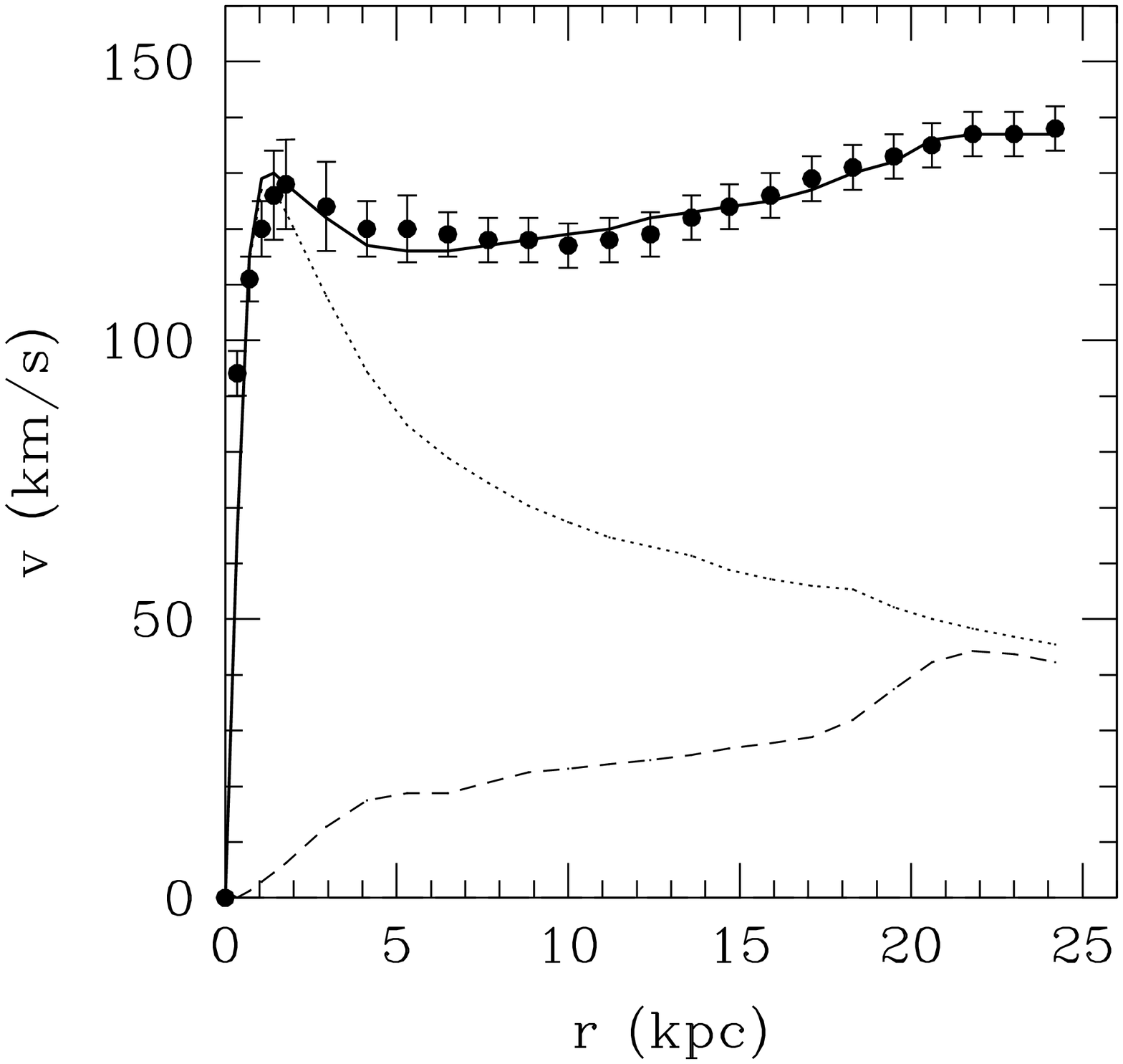}\\
\includegraphics[width=0.65\columnwidth]{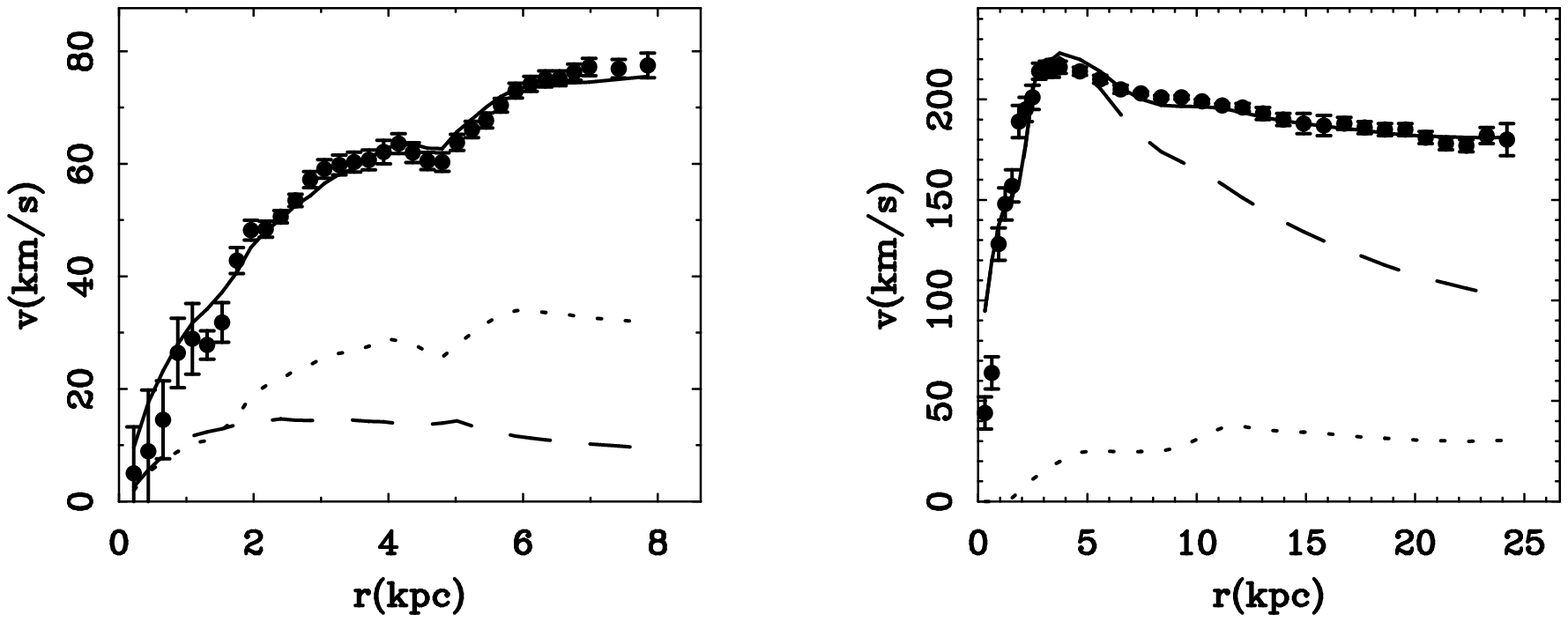}&
\includegraphics[width=0.32\columnwidth]{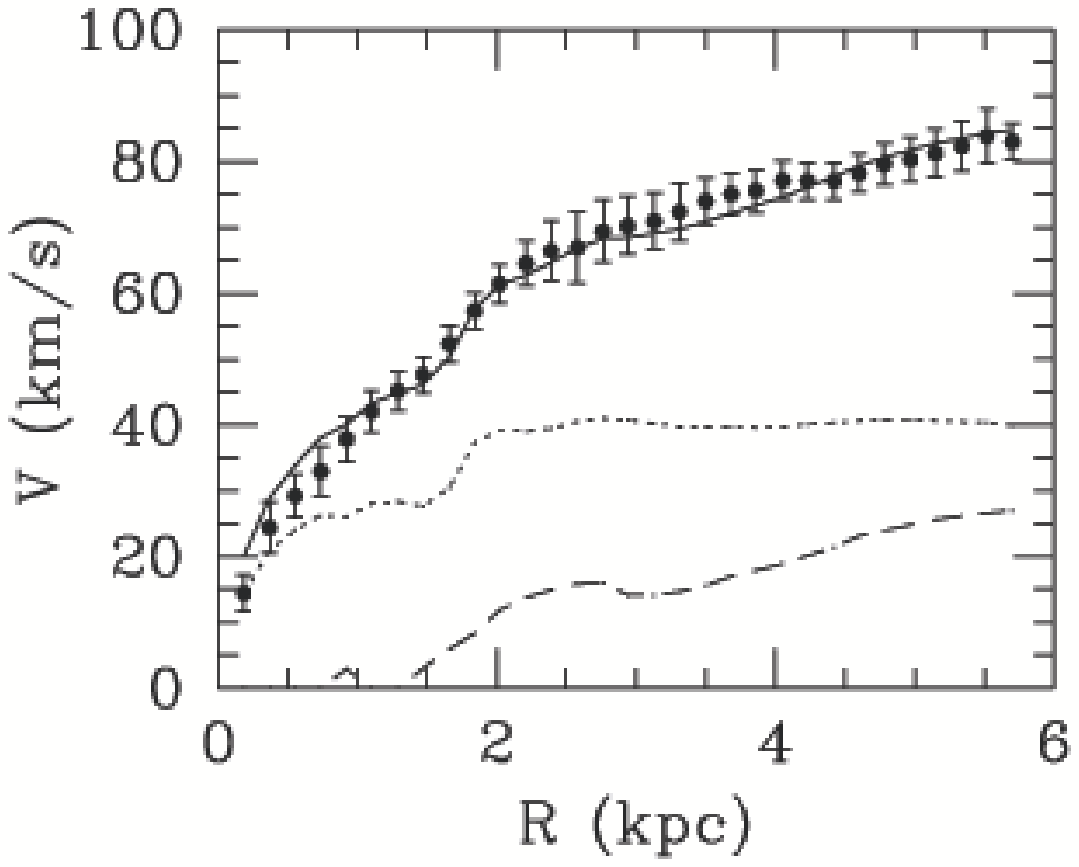}
\end{tabular}
\caption{Observed RCs of five galaxies, from different sources (data points) compared with the MOND predictions (lines going through the data points). Other lines in the figures are the Newtonian curves for various baryonic components.} \label{fig2}
\end{figure}

Over and above the prediction of general galactic laws, MOND predicts the full rotation curves (RCs) of individual galaxies, from only the distribution of their baryonic mass. This is a feat that the DM paradigm cannot perform, and in all probability will never be able to perform. This is because in the DM paradigm, the relation between the baryons and the full potential field of a galaxy (on which the RC depends), strongly depends on the unknowable history of the particular galaxy. In Figure \ref{fig2} are shown the observed RCs of a few disc galaxies of different types, together with the MOND predictions.
\subsection{$\az$?}
$\az$ can be derived in several independent ways from observations of the above laws, and one finds
$\az\approx 1.2\times 10^{-8}~{\rm cm~s^{-2}}$.
Interestingly, it was noticed early on \cite{milgrom83} that $\baz\equiv 2\pi\az\approx c H_0$. And now that a ``cosmological constant'', $\Lambda$, is indicated by the observations, with $\Lambda/3H^2_0\sim 1$, we also note that $\baz\approx c(\Lambda/3)^{1/2}$ might be of significance. This indication that the internal dynamics in galaxies might be related to the state of the universe at large, may turn out to be the most far-reaching implication of MOND. Much has been made of this coincidence in the literature, but this important issue is beyond our present scope.

\subsection{All is not roses}
There are two major conundrums remaining for MOND.
The first is that MOND does not explain away completely the mass discrepancy in galaxy clusters. The mass discrepancy in clusters, which is typically a factor of about 10 for the cluster at large (at a few Mpc from the center) is reduced by MOND to only a factor of about 2-3 (e.g. \cite{sanders99,angus08}). This means that MOND still requires some form of yet undetected matter in clusters. It has been suggested that this could be in the form of difficult-to-detect baryons (a small fraction of the still missing baryons suffice for this), or that it is made of neutrinos.
The remaining mass discrepancy is rather more pronounced in the core, which means that the still-dark matter is rather more centrally concentrated than the observed x-ray emitting gas (it is distributed more similarly to the observed galaxies in the cluster). Such an explanation will also account for the observations of the ``bullet cluster'', which does not present a problem for MOND beyond that already found in individual clusters.

The other conundrum is that we still do not have in MOND a full account of the roles purportedly played by cosmological dark matter. There are several relativistic MOND theories that have not yet been explored in this context. And, in any event, we may have not zeroed in yet on the correct relativistic MOND theory that will account for all aspects of cosmology and structure formation.
\section{MOND theories}

\subsection {Nonrelativistic theories}
Newtonian dynamics of a gravitating system of masses is encapsuled in the law of motion for the masses  ${\bf a}=-\gf$, and the Poisson equation by which the gravitational potential $\f$ is determined from the mass distribution $\r$: $\div(\gf)=4\pi G \r$.

A modified-gravity NR MOND theory is one in which the law of motion remains intact, but the Poisson equation is replaced by a MOND equation. This means that the standard Poisson action is replaced by a new MOND action for the gravitational potential. For years the only such theory known was based on a nonlinear version of the Poisson equation \cite{bm84}
 \beq\div[\mu(|\gf|/ \az)\gf]= 4\pi G\r. \eeqno{gatas}
Recently, a new NR formulation has been proposed \cite{milgrom10}, called QUMOND, in which $\f$ is determined from the two equations
 \beq\Delta\f^*=4\pi G\r,~~~~~~~~\Delta\f=\div[\nu({|\gf^*|/
\az})\gf^*]. \eeqno{mulopa}
It too is derived from an action. Its great advantage is that it requires solving only linear differential equations, even though it describes a fully nonlinear MOND gravity.
\subsection{Relativistic theories}
The first full fledged relativistic MOND theory to be propounded, with proper MOND gravitational lensing, was TeVeS (Tensor-Vector-Scalar Gravity) (\cite{bek04}, building on ideas in \cite{sanders97}). Gravity is described by the fields $ g_{\a\b}, ~{\cal U}_\a,
~\phi$, and matter couples minimally to the ``physical'' metric $\tilde g_{\a\b}=e^{-2\phi}(g_{\a\b} +
{\cal U}_\a {\cal U}_\b) - e^{2\phi} {\cal U}_\a {\cal
U}_\b$. TeVeS is still the most extensively explored MOND theory (see \cite{skordis09} for a review).
With the enhanced interest in relativistic MOND theories brought about by the advent of TeVeS, several new classes of such theories have been proposed. I list them briefly, with an emphasis on my own hobby horse--bimetric MOND theories.

In MOND adaptations of Aether theories \cite{zlosnik07}, a vector field of unit length is added whose free action is an appropriate MOND function of
scalars quadratic in the vector-field derivatives.

Galileon k-mouflage MOND adaptations \cite{babichev11} are, like TeVeS,  tensor-vector-scalar theories. They are said to improve on TeVeS in various regards (e.g., small enough departures from GR in high-acceleration environments, thus strongly conforming to solar-system and binary-pulsar constraints).

Nonlocal metric MOND theories \cite{soussa03,deffayet11} are
pure metric, but highly nonlocal in that they involve operators that are functions of the 4-Laplacian.
\subsection{Bimetric MOND theories}
 In bimetric MOND (BIMOND) theories \cite{milgrom09,milgrom10a,milgrom10b}
gravity is described by two metrics $\gmn$ and $\hgmn$. Matter couples only to $\gmn$, but there possibly exists ``twin matter'' (TM) that couples only to $\hgmn$ in a similar way. The two metrics are governed each by a standard Einstein-Hilbert action; but the crux of the theory is an interaction between the metrics, which engenders all the MOND effects.
It is a function of scalars constructed from
 $\az^{-1}C\ten{\a}{\b\c}$, where
 $C\ten{\a}{\b\c}=\C\ten{\a}{\b\c}-\hC\ten{\a}{\b\c}$. In particular, I found the choice of scalars $\Up=\Gmn\Up_{\m\n},~~~\hat\Up= \hGmn\Up_{\m\n}$, based on the tensor $\Up_{\m\n}=
C\ten{\c}{\m\l}C\ten{\l}{\n\c} -C\ten{\c}{\m\n}C\ten{\l}{\l\c}$, particularly felicitous.

The most appealing versions of BIMOND have the following properties: In the limit $\az\rar 0$ they go to GR with a cosmological constant of order $\az^2$, according with the above mentioned ``coincidence''. In the NR, slow-motion limit, gravity is described by two potentials $\f$ and $\fh$, with the metrics
  \beq\gmn=\emn-2\f\d_{\m\n},~~~~
 \hat g_{\m\n}=\emn-2\fh\d_{\m\n}. \eeqno{tamana}
This implies that lensing occurs as in GR, with photons and massive particles ``seeing'' the same potential.
Matter senses only $\f$, and TM only $\fh$.
The potentials are determined from
 $\f=\ft/2+\fb$, $\fh=\ft/2-\fb$ with
  \beq\D\ft=\fpg(\r+\rh),~~~~~
 \div\{\mt(|\gfb|/\az)\gfb\}=\fpg(\r-\rh), \eeqno{satara}
$\r$ being the density of matter and $\rh$ that of TM. Matter and TM do not interact in the high-acceleration regime. In the deep-MOND regime they behave exactly as if they had opposite signs of the gravitational mass: they repel each other, while each attracts its own kind.
\par
There are also interesting ``microscopic'' approaches to MOND, which are listed, with references, in \cite{milgrom11}.
\section{Modified inertia}
All the MOND theories described above are of the MG type: They involve modification of only the action of the gravitational field (the NR, Poisson action, or the relativistic, Einstein-Hilbert action for the metric).
A different possible approach to MOND is to modify the kinetic action of particles, or, more generally, in the relativistic case, modify the matter action. In the NR case, one way to do this for a system of gravitating particles, is to leave the Poisson
equation intact, and to modify the particle equation of motion into
$\AA[\{\vr(t)\},t,\az]=-\gf[\vr(t)],$ instead of the Newtonian $\va=-\gf$. Here $\AA$ is a functional of the full particle trajectory with the dimensions of acceleration.
More generally, for a test particle of mass
$m$, subject to a force $\vF(t)$, a model equation of motion would be
 \beq m\AA[\{\vr(t)\},t,\az]=\vF(t).  \eeqno{jutreq}
Universality of free fall is then guaranteed. The standard limit is $\AA[\{\vr(t)\},t,\az\rar 0]\rar\va$, and scale invariance in the deep-MOND limit dictates
$\vA[\{\vr(t)\},t,\az\rar\infty]\rar \az^{-1}\vq[\{\vr(t)\},t]$.
\par
This direction has proven less tractable, and remains relatively unexplored (for existing studies see \cite{milgrom94,milgrom99,milgrom02,milgrom06}). I think, however, that it is quite attractive, and definitely worth pursuing: It is a natural framework for connecting MOND with cosmology, as
perhaps pointed to by the above coincidence. In its nature it calls for rethinking the whole concept of inertia. Modifying inertia is not a new concept, or one unique to MOND. For example, special relativity replaces $\va$ in Newton's second law by $\AA=d(\c\vv)/dt=\c\va+\c^3(\vv\cdot\va)\vv$. There are also many known instances in physics where inertia is created, or strongly modified, as an effective result of the interaction of the particle with a medium (e.g., electrons in solids, or photons in a refractive medium); MI, or emergent inertia, is the theme of the Mach principle. MOND MI could have come about in a similar way. Some suggestions in this vein are described, e.g., in \cite{milgrom99,pikhitsa10,klinkhamer11}.
\par
But, my purpose is not to consider MI theories {\it qua} theories; in fact, I cannot, at present, offer any new such theory. Here, I am trying to anticipate some of the major differences in predictions between MI and MG.
\par
Two important results emerged from the detail consideration of MOND MI, in \cite{milgrom94}: The first is phenomenological: In any MI theory the relation between the central gravity force, the orbital radius, and the orbital speed on circular orbits (describing RCs of disc galaxies) perforce has the form
\beq \frac{V^2}{R}\mu\left(\frac{V^2}{R\az}\right)=-{\partial \f\over\partial
R},\eeqno{mushta}
where the ``interpolating function'', $\mu(x)$, is universal for the theory. This is easily seen to follow from eq.(\ref{jutreq}) applied to circular orbits.
$\mu$ was derived in \cite{milgrom94} in terms of the action of the theory. Note that it is specific to circular orbits, and is not necessarily relevant to other orbits.
\par
The other result is that a (NR) MI MOND theory that is derived from an action, and that has Galilei invariance must be non local in time (which means that the action cannot be a time integral of a Lagrangian that depends on a finite number of derivatives).
Add to this the inherent nonlinearity of MOND and we see why it has been difficult to construct such theories, and why  even toy, heuristic models are rather unwieldy. All this does not, however, by any means, argue against this approach. For example, the general result (\ref{mushta}), the like of which does not exist in MG theories, may testify to the usefulness of the approach.
\subsection{Toy models and their consequences}
Some progress can be made by considering heuristic models that will, at least, point to what we can expect from such theories. In particular, they provide warnings that we should not embrace all MOND predictions of MG theories as inescapable predictions of MOND at large.
For example, in \cite{milgrom94} I considered the kinetic action for a unit mass particle of the form
 \beq \sk=\lim_{T\rar\infty}{1\over 4 T}\int_{-T}^T\vv\cdot
 F\left({\sk \DD\over\azs}\right)\vv~dt, \eeqno{zzii}
with $F(x\gg 1)\approx 1$, $F(x\ll 1)\propto x^{1/3}$; $\DD$ indicates time derivatives acting on both sides. The multi-dimensional, anisotropic harmonic oscillator was solved exactly, and point to interesting interplay between the motions along the different axes. But the theory is difficult to solve for other problems.
\par
Another class of toy models can be constructed by working in Fourier space, which naturally accounts for nonlocality: The Newtonian equation for a unit-mass test particle, $\va_N(t)=\vf(t)$, which in Fourier space  reads $ \haNo=\hfo$ (hatted quantities are Fourier components)
is replaced by
 \beq \hao
 \k[\Ao/\az]=\hfo=\haNo,\eeqno{iiib}
where $\Ao$ is a real, positive functional of the trajectory, and
possibly a function of $\o$, with the dimensions of acceleration.
As is typical of MI theories (e.g., SR) this produces an acceleration $\va(t)$ that is not, in general, in
the direction of $\vf(t)$. For Galilei invariance, construct $\Ao$ from $\hao$. Under time translations, $\hao\rar e^{i\o t_0}\hao$; so, construct $\Ao$ out of $|\hao|$.
\par
Such models are surely primitive. They are not guaranteed an action underpinnings, and may lack other desiderata (such as causality). There is also no indication that MOND is so describable in frequency space. But such models have a heuristic value similar to the pristine formulation of MOND in \cite{milgrom83}, as a simple relation between momentary force and acceleration.
\par
The simplest choice is to take a frequency-by-frequency MOND modification, with $\Ao=|\o\hao|$, and the appropriate form of $\k$. Such a model will make all the salient MOND predictions that are not concerned with coupling between different motions: predictions 1-6 in the list of Kepler-like laws above, and, more generally, the correct rotation-curve predictions. But prediction 7, while still qualitatively valid, will differ greatly from that of MG versions: the disc discrepancy here depends on the vertical accelerations, not on the rotational ones, since the two motions are typically of very different frequencies. Also, such a theory will predict no EFE, unless the external and internal motions are of similar frequencies.
It will, on the other hand, predict the observationally correct center-of-mass motions (CoMMs) of bodies with disparate values of the internal and external accelerations (such as stars with high internal accelerations moving MOND-like in a low acceleration field of a galaxy): because of the disparate frequencies, each motion is modified according to its own characteristic acceleration.
\par
I think that an EFE in MOND is observationally required, so let us make $\Ao$ depend not only on $\o$, but on all components with frequency up to $\o$.
For example (heuristically again) take
\beq \Ao\propto\int_{-|\o|}^{|\o|}|\hat
\va(\o')|d\o'\propto\int_{0}^{|\o|}|\hat \va(\o')|d\o', \eeqno{v}
where I use the fact that $|\hao|$ is symmetric, since $\hat
\va^*(\o)=\hat \va(-\o)$.
\par
Some of the consequences of such a theory, beyond laws 1-6, and the full MOND prediction of RCs, eq.(\ref{mushta}), are:
MOND correction to vertical motions in disc galaxies is determined typically, by the acceleration of the rotational motion, which is of a higher acceleration, but lower frequency. The CoMM of bodies with high internal acceleration is correct (MOND-like), because the internal motions are perforce at higher frequencies, and so do not affect the low-frequency CoMM ($\o_{in}/\o_{out}\sim [(a_{in}/a_{out})(r_{out}/r_{in})]^{1/2}\gg 1$).
An EFE is predicted, but its implications might be very different than in MG theories. Take, e.g., a dwarf spheroidal, or a globular cluster, on orbit in the field of a galaxy. In MG theories the EFE on the internal dynamics depends only on the field of the mother galaxy at the present position of the subsystem. In the above model, the EFE is dominated by the orbital component with the largest acceleration, as long as its frequency is smaller than the intrinsic ones, in a hysteresis effect typical of time nonlocal theories. So the EFE might be dominated, e.g., by the galaxy's field at orbital perigalacton.
\par
High-acceleration Fourier components (having $|\va(\o)|\o\gg \az$), such as for the planetary motions around the sun, suffer only a small MOND correction [negligibly small if $\k(x)$ approaches $1$ very fast as $x\gg 1$]. But for bodies whose orbit takes them far from the sun, to regions where the accelerations are small;
namely, orbits with Fourier component for which $|\va(\o)|\o\ll \az$--such as very-long-period comets--a low frequency anomalous acceleration of order $\az$ may appear with that frequency, even when the body is in a high acceleration region, e.g., near the sun.
\par
This is only a very brief and superficial look at what may lie in the domain of MI theories. I hope it is enough to arouse more interest in them, on one hand, and on the other to warn against adopting all the predictions of existing MOND theories, beyond the basic ones, as absolute predictions of the MOND paradigm.

\clearpage

\end{document}